# Reaction Cycles of Halogen Species in the Immune Defense: Implications for Human Health and Diseases and the Pathology and Treatment of COVID-19

**Qing-Bin Lu***

Department of Physics and Astronomy and Department of Biology, University of Waterloo, 200 University Avenue West, Waterloo, Ontario, Canada
*Correspondence: qblu@uwaterloo.ca

**Abstract:** There is no vaccine or specific antiviral treatment for COVID-19 that is causing a global pandemic. One current focus is drug repurposing research, but those drugs have limited therapeutic efficacies and known adverse effects. The pathology of COVID-19 is essentially unknow. Without this understanding, it is challenging to discover a successful treatment to be approved for clinical use. This paper addresses several key biological processes of reactive oxygen, halogen and nitrogen species (ROS, RHS and RNS) that play crucial physiological roles in organisms from plants to humans. These include why superoxide dismutases, the enzymes to catalyze the formation of $H_2O_2$, are required for protecting ROS-induced injury in cell metabolism, why the amount of ROS/RNS produced by ionizing radiation at clinically relevant doses is ~1000 fold lower than the endogenous ROS/RNS level routinely produced in the cell, and why a low level of endogenous RHS plays a crucial role in phagocytosis for the immune defense. Here we propose a plausible amplification mechanism in the immune defense: ozone-depleting-like halogen cyclic reactions enhancing RHS effects are responsible for all the mentioned physiological functions, which are activated by $H_2O_2$ and deactivated by NO signaling molecule. Our results show that the reaction cycles can be repeated thousands of times and amplify the RHS pathogen-killing (defense) effects by 100000 fold in phagocytosis, resembling the cyclic ozone-depleting reactions in the stratosphere. It is unraveled that $H_2O_2$ is a required protective signaling molecule (angel) in the defense system for human health and its dysfunction can cause many diseases or conditions such as autoimmune disorders, aging and cancer. We also identify a class of potent drugs for effective treatment of invading pathogens such as HIV and SARS-CoV-2 (COVID-19), cancer and other diseases, and provide a molecular mechanism of action of the drugs or candidates.

**Keywords:** respiratory burst; phagocytosis; reactive oxygen species; reactive halogen species; reactive nitrogen species; signaling; antitumor agents; antiviral agents; HIV; COVID-19

## 1. Introduction

*Reactive oxygen species* (ROS), including superoxide ($O_2^{\bullet-}$), hydrogen peroxide ($H_2O_2$) and hydroxyl radical ($OH^{\bullet}$), are constantly produced from aerobic metabolism of the cell. It is known that $OH^{\bullet}$ is potent but non-selective oxidant, while $O_2^{\bullet-}$ and $H_2O_2$ are far less reactive. Moreover, increasing evidence shows that $H_2O_2$ is a necessary signaling molecule in the regulation of a variety of biological processes such as cell proliferation and differentiation, tissue repair, immune cell activation, circadian rhythm, vascular remodeling and aging [1-5]. $H_2O_2$ is also a signaling molecule of plant defense against pathogens [6].

Rising evidence has also clearly demonstrated the critical physiological role of superoxide dismutases (SOD) in humans and all other mammals, which are the enzymes converting $O_2^{\bullet-}$ into $H_2O_2$) [7]. For instance, it was observed that genetic knockout of SOD produced deleterious phenotypes in organisms ranging from bacteria to mice [8-11]. Knockout mice lacking Mn-SOD died soon after birth or suffered severe neurodegeneration [8]. Mn-SOD was also shown to be required for protecting ROS-induced injury in $O_2$ metabolism [11]. Mice lacking cytosolic Cu,Zn-SOD developed multiple pathologies, including liver cancer [9], muscle atrophy, cataracts, and a reduced lifespan [10]. Research on biological action of ionizing radiation (IR) also showed that modifying SOD activity in bacteria reduced the oxygen radiosensitizing effect and that SOD injection following whole body x-irradiation significantly protected mice from lethality [12,13]. Also, active SOD or SOD mimic compounds inhibited IR-induced deleterious effects including transformation assays, bystander effects, normal tissue damage associated with inflammatory responses and fibrosis [14,15]. These studies clearly demonstrated that $O_2^{\bullet-}$ was somehow participating in radiation-induced injury, whereas $H_2O_2$ (SOD) was required for normal physiological functions. Given that $O_2^{\bullet-}$ is a weak oxidant and the downstream formation of $OH^{\bullet}$ from $H_2O_2$ is responsible for causing oxidative damage, the precise role of $H_2O_2$ (or $O_2^{\bullet-}$) in these critical physiological functions is essentially unknown [15].

*Respiratory burst* is the rapid release of reactive species from different types of cells, particularly myeloid cells (phagocytes) such as neutrophils, monocytes, macrophages, mast cells, dendritic cells, osteoclasts and eosinophils. It is usually utilised for immunological defence. Exposure to ROS or *reactive halogen species* (RHS) or *reactive nitrogen species* (RNS) kills the engulfed pathogens, resulting in pathology of infection. Similar 'respiratory burst' is found when cells are exposed to IR at a clinical radiation dose [15] or when the ozone layer is exposed to ionizing cosmic rays in the polar stratosphere especially in the presence of halogenated molecules (particularly freons, CFCs) [16-19].

*Phagocytosis* is the process by which a cell uses its plasma membrane to engulf a pathogen, forming an organelle called phagosome. It plays a major role in the *innate immune defense* by effectively killing of pathogens. The phagosome then moves toward the centrosome of the phagocyte and fuses with lysosomes, forming a phagolysosome for degradation of pathogens. A phagocyte has various receptors on its surface, including opsonin receptors, scavenger receptors and Toll-like receptors. The widely accepted belief is that phagocytosis is most effective for killing microorganisms (e.g., bacteria), while the *adaptive immune defense*, mainly by lymphocytes, is more important for viral infections. However, increasing evidence has shown that phagocytosis plays a crucial role in killing a variety of intracellular or cytoplasmic viruses ranging from influenza, mumps, measles and rhinovirus to Ebola and HIV, either directly protecting against viral infections in the innate immune defense [20-25] or through the antibody-dependent cellular phagocytosis providing mechanisms for clearance of virus and virus-infected cells and for stimulation of downstream adaptive immune responses by antigen presentation or the secretion of inflammatory mediators [26-28]. Particularly in humoral (adaptive) immunity, B cells upon activation produce antibodies, each of which recognizes a unique antigen and neutralizes a specific pathogen. The binding of antibodies to antigens makes the latter easier targets for phagocytes and gives rise to several protective mechanisms. These include the activation of complement to cause inflammation and cell lysis, the opsonization by coating antigen with antibody (e.g., immunoglobulin G, IgG) or complement to enhance the phagocytosis of pathogens through binding to the opsonin receptors of phagocytes, and antibody-dependent cell-mediated cytotoxicity by attaching antibodies to targeted cells to cause the latter's destruction by macrophages, eosinophils, and natural killer (NK) cells. Interestingly, ROS-regulated RHS formed by phagocytes showed the strong viricidal effect on HIV-1 [21, 22].

The importance of bats in the evolution of human SARS-CoV-2 that causes the disease COVID-19 has been well documented and indirectly supported by a high degree of similarity with the bat coronavirus genomes. High levels of ROS in bats have been associated with the negative effect on the proof-reading activity of coronavirus RNA polymerases, and it was hypothesized that coronavirus infection in bats is mostly asymptomatic (i.e., bats maintain and disseminate deadly viruses) because high levels of ROS can indirectly enable the bats to control the levels of viral replication [29]. The bat genes most likely to reflect this phenomenon are those directly related to the first line of antiviral defense—the innate immune system [29].

$O_2$-dependent degradation in phagocytosis depends the production of ROS, RHS and RNS, as shown in Figure 1. First, the NADPH oxidase (NOX2) embedded in the phagolysosome membrane is activated to produce $O_2^{\bullet-}$ by $O_2$ capturing an electron from cytosolic NADPH. $H_2O_2$ is generated from $O_2^{\bullet-}$ via superoxide dismutases (SOD) or heme(halo)peroxidases (MPO, EPO and LPO); $OH^{\bullet}$ is then generated via the Haber-Weiss reaction. *Through MPO/EPO/LPO, $H_2O_2$ activates a halogenating system, which generates hypochlorous acid (HOCl) and kills pathogens* [30-34]. The inducible nitric oxide synthases (iNOS) also catalyze the production of $NO^{\bullet}$ from L-arginine; $NO^{\bullet}$ then reacts with $O_2^{\bullet-}$ to produce a peroxynitrite ($ONOO^-$) that forms $NO_2$ (via reaction with $CO_2$) [35-37]. The last reaction competes obviously with the SOD-catalyzed scavenging of $O_2^{\bullet-}$.

All $OH^{\bullet}$, HOCl and $ONOO^-$ ($NO_2$) are strong oxidants believed to damage DNA/RNA, lipids and proteins, and have anti-pathogen effects [30-37]. Particularly HOCl has widely been used as a strong disinfectant; HOBr and HOI at lower levels were also detected in stimulated immune cells [30-34]. Macrophages have weaker antiviral properties than neutrophils and their activation is made by transient respiratory burst, which regulates the inflammatory response by inducing synthesis of cytokines for redox signaling. Due to the high toxicity of ROS/RHS/RNS, damage to host tissues can also occur during inflammation.

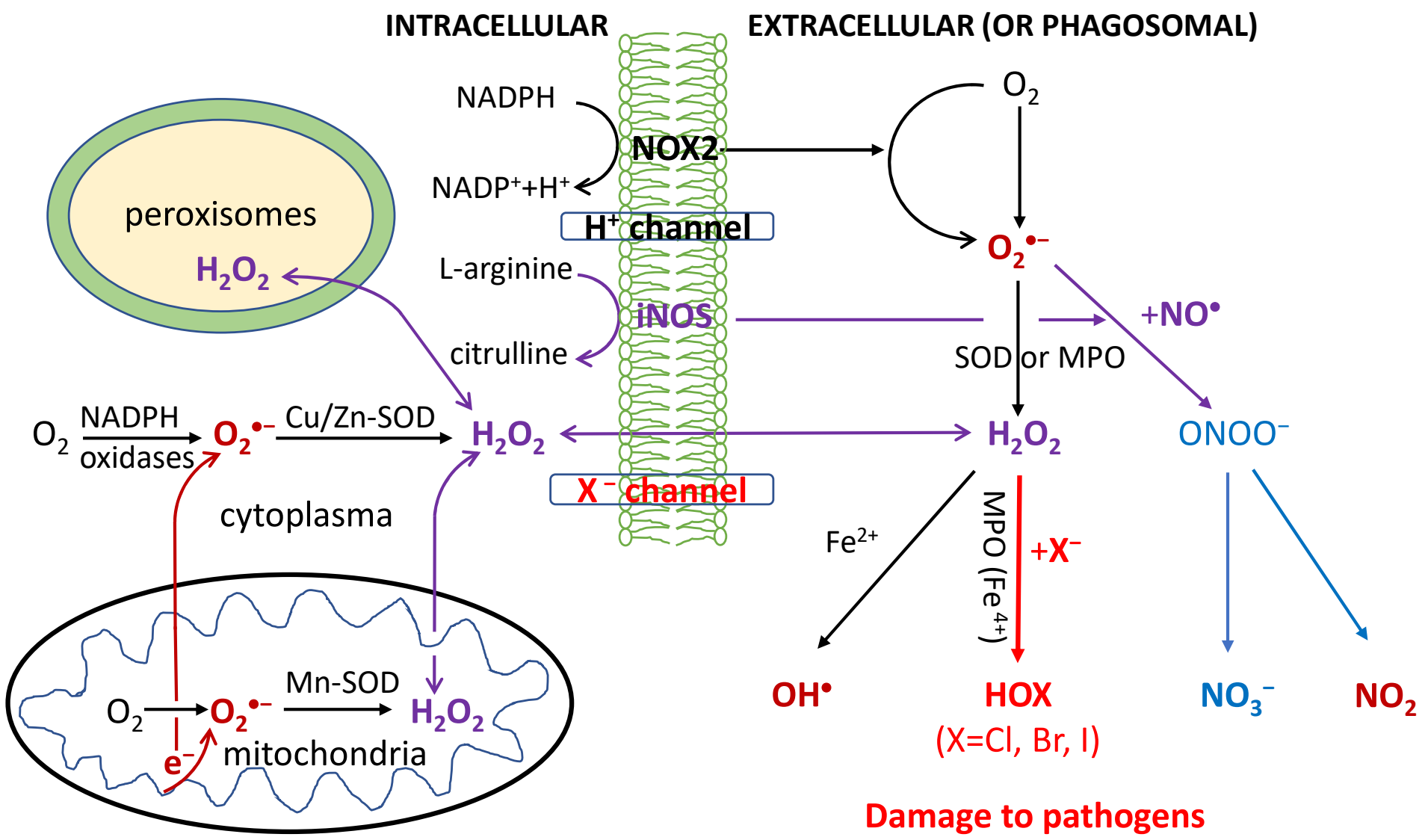

**Figure 1.** Respiratory burst in the immune defense against pathogens in phagocytosis. Reactive oxygen/halogen/nitrogen species ($OH^{\bullet}$; HOX with X=Cl, Br or I; $NO_2$) produced within the phagosome cause damage to the DNA/RNA/lipids/proteins of phagocytosed pathogens. The formyl peptide receptors (FPRs) stimulate the activation of the phagocytic NADPH oxidase (NOX2) to form $O_2^{\bullet-}$, which is rapidly converted to $H_2O_2$ by the action of cytoplasmic or mitochondrial SOD enzymes. Additionally, $H_2O_2$ can also be produced extracellularly by the IgG-catalyzed oxidation of water or by the interactions of other receptors such as tumor necrosis factor (TNF) receptors, Toll-like receptors (TLRs) and Nod-like receptors (NLRs) with pathogen-associated ligands. The ROS are also produced by a variety of cell types (nonphagocytic cells) and tissues, catalyzed by six other members of the NOX family including nonphagocytic enzymes NOX1, NOX3, NOX4, NOX5, and the dual oxidases Duox1 and Duox2. The activation of these NOX are stimulated by growth factors, cytokines, and integrins. NOX5 and Duox1 and Duox2 are activated by $Ca^{2+}$ through a cytoplasmic calcium-binding domain. $H_2O_2$ and $NO^•$ can diffuse freely across membranes as indicated by the arrows.

Of particular interest is the role of *endogenous RHS* in the $O_2$-dependent killing mechanisms of invading pathogens in phagosomes. It is worthy of mention that HOXs (X=Cl, Br) are also important halogen species causing $O_3$ depletion in the Earth's stratosphere [38,39]. They are derived from solar photolysis [40] or cosmic-ray driven *dissociative-electron-transfer (DET)* reactions [16-19] of halogenated organic molecules, especially CFCs. In radiobiology, there is also a long mystery that *the amount of primary ROS produced by IR at clinically relevant radiation doses (2Gy) is ~1000 fold lower than the background (endogenous) ROS level produced by oxidative metabolism of the cell*, as first noted by Ward [41]. This leads to the question of how the ROS produced by the few primary ionization events is amplified to account for observed biological effects. *The mechanism remains elusive, as recently reviewed* [15]. This vividly reminds the hypothesis of significant CFC-induced ozone loss in the Earth's stratosphere by atmospheric scientists in the 1970s, giving birth of modern ozone depletion theory. It is now well established that there is an amplifying mechanism where a Cl atom catalytically destroys up to 100000 ozone molecules via halogen reaction cycles [38,39]. Note that during respiratory burst, the increases in $O_2$ (not $NO^•$) consumption were typically observed to be only 2-20 fold among different cells and/or conditions [42]. *This, together with the mysterious critical physiological role of $H_2O_2$, implies that a drastic amplification mechanism is clearly required.*

## 2. Theoretical Studies

### 2.1. Halogen Cyclic Reactions

*Here we hypothesize* that (i) the *initial* yields of endogenous ROS/RHS/RNS produced in the phagosome are not sufficient to cause a significant anti-pathogen effect in the immune defense, for which an amplification mechanism is required; (ii) The $O_3$-depleting-like reaction cycles of RHS take place in the phagosome for immunological defences in cells of humans and all other mammals, via enzymes such as hemeperoxidases or other antioxidant enzymes, instead of solar photons or cosmic rays; (iii) Similar to the formation of polar $O_3$ hole, numerous reaction cycles of RHS (mainly $X^•$, $XO^•$, HOX, $OX^-$, $X_2$; X=Cl or Br or I) play a major role in killing pathogens and immunological defences. These are schematically shown in Figure 2.

In biological systems, the halogenation of organic molecules is catalyzed by enzymes haloperoxidases such as MPO, EPO and LPO in normal physiological processes, which combine the inorganic substrates $X^-$ and $H_2O_2$ to produce RHS. RHS in turn oxidize the hydro-carbon or -nitrogen substrate RH/RNH to synthesize many halogenated organic compounds (RX/RNX):

Biosynthesis: $X^- + H_2O_2 \rightarrow HOX/OX^-/OX^•/X_2 \rightarrow RX/RNX$ (1)

DET reaction: $RX/RNX + e^- \rightarrow RX^-/RNX^- \rightarrow (X^• + R^-/RN^-)$ or $(X^- + R^•/RN^•)$ (2)

By reaction (1), many halogenated organic compounds are biosynthesized [32,34,43, 44]. It is also well known that halogenated RX/RNX are highly effective for DET reactions that result in anions $R^-/RN^-/X^-$ or reactive radicals $X^•/OX^•/R^•/RN^•$, as expressed in reaction (2) [15-19, 45-51]. As shown in Table 1, all of these RHS have been found in biological systems especially phagosomes [30-34,43, 44,52-54]. One should be cautious with some reaction rate constants listed in Table 1 as they are somewhat outdated. For example, most experiments by nanosecond- or picosecond-resolved radiolysis of aqueous halogenated molecules actually measured the DET with the hydrated electron ($e_{aq}^-$) rather than the subpicosecond-lived prehydrated electron ($e_{pre}^-$). $e_{aq}^-$ is deeply bound at ~3.5 eV and is relatively inert, while $e_{pre}^-$ has a small binding energy of 1.0-1.5 eV and is extremely effective for DET with halogenated molecules, as well demonstrated and reviewed previously [49-51,55,56,18,19].

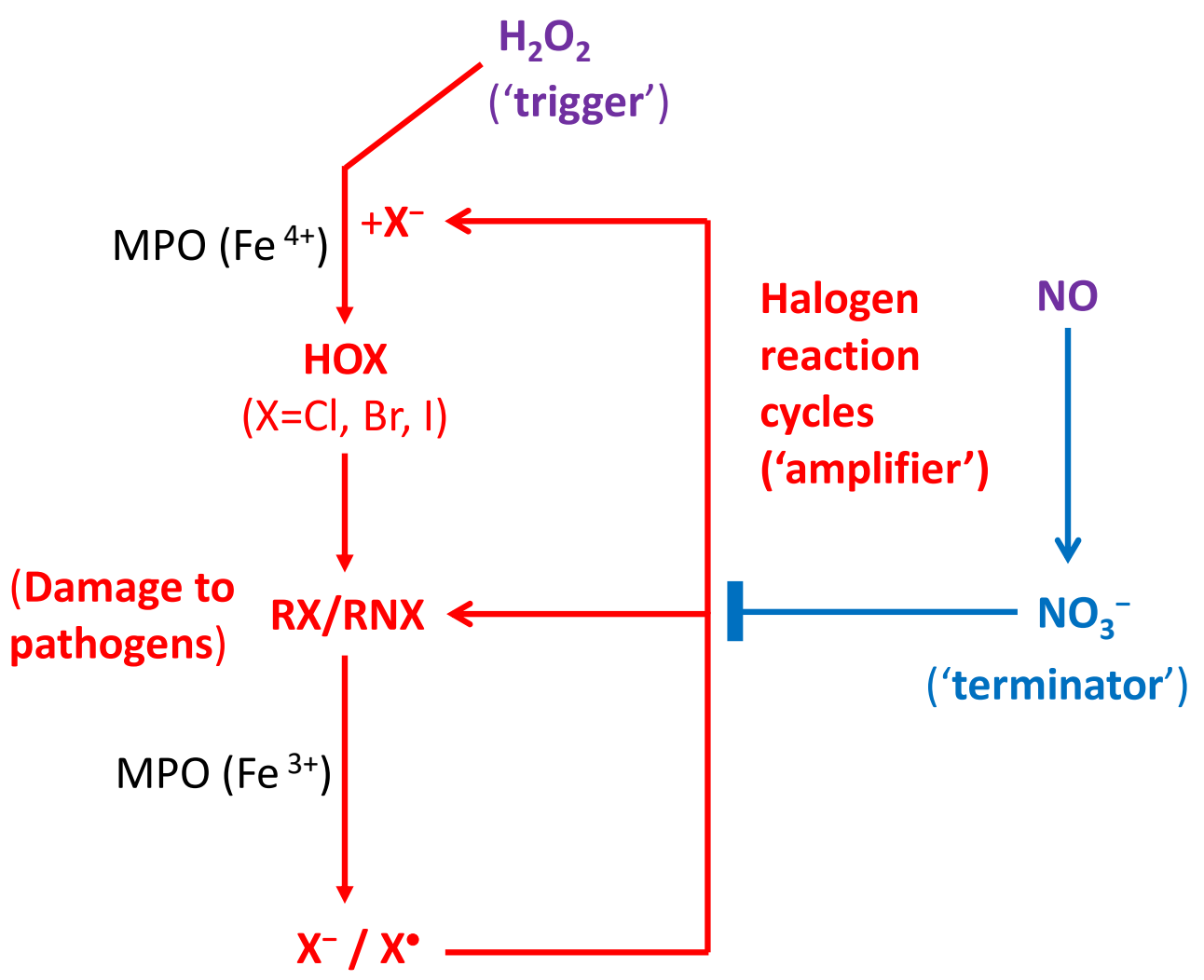

**Figure 2.** Proposed halogen reaction cycles involved in the immune defense against pathogens and in the inflammatory process. The amplified reactions of reactive halogen species (RHS), including HOX, $OX^-$, $X_2$, $X^•$ and $OX^•$ (X=Cl, Br or I), produced by dissociative electron transfer (DET) within the phagosome cause major damage (represented by RX/RNX) to the DNA/RNA/lipids/proteins of phagocytosed pathogens. The immune defence system consists mainly of three components: the $H_2O_2$ signaling pathway as the trigger, the RHS reaction cycles as the amplifier, and the $NO^•$ signaling pathway as the terminator. $H_2O_2$ from other intracellular organelles such as mitochondria and peroxisomes can also diffuse across membranes and promote the anti-pathogen activity.

**Table 1.** Approximate rate constants ($M^{-1}$ $s^{-1}$) for aqueous-phase reactions of some species involved in phagocytosis at pH=7.0, 288K.

| Reaction | Products | *k* ($M^{-1}$ $s^{-1}$) | Refs |
|---|---|---|---|
| $X^- + H_2O_2 + MPO + H^+$<br>$X^- + H_2O_2 + EPO + H^+$ | $HOX/OX^-/X_2 + H_2O$ | $2.5\times10^4$, $1.1\times10^6$, $7.2\times10^6$ (X=Cl, Br, I)<br>$3.1\times10^3$, $1.9\times10^7$, $9.3\times10^7$ (X=Cl, Br, I) | [31] |
| $Cl^- + OH^• + H^+$ | $Cl^• + H_2O$ | $1.5\times10^{10}$ | [39] |
| $R_2NH + HOX/XO^-$ | $R_2NX$ | $10^7$-$10^8$ | [44] |
| $Cl_2 + RNH_2$ | $RNHX + Cl^- + H^+$ | (>)$10^9$ | [44] |
| $NH_2Cl + e_{aq}^-$ | $Cl^- + NH_2$ | $2.2\times10^{10}$ | [53] |
| $R_2NCl + e_{aq}^-$ | $Cl^- + R_2N^•$ | $1.5$-$1.9\times10^{10}$ | [52,53] |
| $RNHCl + e_{aq}^-$ | $Cl^- + RNH^•$ | $6.1$-$9.3\times10^9$ | [52,53] |
| $RR'NBr + e_{aq}^-$ | $Br^• + RR'N^-$ | $2.9\times10^{10}$ ($1.1\times10^{11}$) | [52,53] |
| $RR'NBr + O_2^-$ | $Br^• + RR'N^-$ | $2$-$9\times10^8$ | [52,53] |
| $NO_3^- + e_{pre}^-$ | $NO_3^{2-}$ | $1.2\times10^{13}$ | [55,56] |
| $CO_3^- + O_2^{•-}$ | $CO_3^{2-} + O_2$ | $4\times10^8$ | [39] |
| $NO_2 + OH^•$ | $NO_3^- + H^+$ | $1.3\times10^9$ | [39] |

Reactions (1) and (2) lay a solid foundation for halogen cyclic reactions to be presented below. Notably, the standard reduction potentials at pH 7 of all redox couples (Compound I/native enzyme, $H_2O$) of enzymes MPO/EPO/LPO abundant in phagosomes are relatively small, being 1.09-1.16 V [31, 32]. Similarly, the standard reduction potentials at pH 7 of the redox couple ($HOX/X^-$, $H_2O$) for two-electron oxidation of halides are 0.78, 1.13 and 1.28 V for X=I, Br and Cl, respectively [31, 32]. Moreover, the exothermic energies for the DET reaction in water of halogenated organic compounds such as RNX is $E_A(X)$-D(N-X)+$E_P$ [45,46,49-51], where $E_A(X)$ is the electron affinity of a X atom [$E_A(Cl)$=3.61 eV; $E_A(Br)$=3.36 eV; $E_A(I)$=3.06 eV] or the neutral radical $RN^•$ [$E_A(RN)$=~3.46 eV] [57], D(N-X) is the known bond dissociation energy of N-X [D(N-Cl)=2.08 eV; D(N-Br)=2.50 eV; D(N-I)=1.65 eV], and $E_P$ ≈1.30 eV [49,50] (1.25-1.50 eV) [58] is the polarization energy of an anon in water. The thus calculated exothermicities for the DET reactions of RNX to form a $RN^•$ and a $X^-$ (or a $RN^-$ and a $X^•$) are 2.83 (or 2.68), 2.16 (or 2.26) and 2.71 (or 3.11) eV for X=Cl, Br and I, respectively, significantly larger than the reduction energies of the redox couples (Compound I/native enzyme, $H_2O$) of enzymes MPO/EPO/LPO. For X=Cl, the DET of RNX favors the dissociation pathway to form $RN^{\bullet}$ (and $X^-$), whereas it preferentially leads to the production of $X^{\bullet}$ (and $RN^-$) for X=Br and I. These estimates indicate that the catalyzed cyclic reactions of RHS must *effectively* take place and sustain for numerous cycles in phagosomes with those endogenous electron-donating enzymes replacing solar photons or cosmic-ray-produced electrons. This is analogous to the established $O_3$ depleting cyclic reactions. For example, $X^{\bullet}$ and HOX with X=Cl, Br, or I can have the following cyclic reactions (**bold** represents altered targets such as damaged DNA/RNA/proteins in pathogens):

$$X^{\bullet} + R_1NH \rightarrow \mathbf{R_1NX} + H^+ \quad (3a)$$

$$\mathbf{R_1NX} + MPO(Fe^{3+}) \rightarrow \mathbf{R_1NX^-} + MPO(Fe^{4+}) \quad (3b)$$

$$\mathbf{R_1NX^-} \rightarrow \mathbf{R_1N^-} + X^{\bullet} \quad (3c)$$

$$\mathbf{R_1NX^-} \rightarrow X^- + \mathbf{R_1N^{\bullet}} \quad (3d)$$

$$\mathbf{R_1N^{\bullet}} + R_2NH \rightarrow \mathbf{R_1NR_2NH} \quad (3e)$$

$$X^- + H_2O_2 + MPO(Fe^{4+}) \rightarrow \ldots \rightarrow X^{\bullet} + MPO(Fe^{3+}) \quad \text{[see reactions (1) \& (2)]} \quad (3f)$$

$$X^{\bullet} + R_2NH \rightarrow \mathbf{R_2NX} + H^+ \quad \text{(continue from 3c)} \quad (3g)$$

$$\mathbf{R_2NX} + MPO(Fe^{3+}) \rightarrow MPO\ (Fe^{4+}) + \mathbf{R_2NX^-} \quad (3h)$$

$$\mathbf{R_2NX^-} \rightarrow \mathbf{R_2N^-} + X^{\bullet} \quad (3i)$$

$$X^{\bullet} + R_3NH \rightarrow \mathbf{R_3NX\ldots} \rightarrow X^{\bullet} \text{ (multiple cycles)} \quad (3j)$$

Net: $H_2O_2+R_1NH+R_2NH+R_3NH \rightarrow \mathbf{R_1NX+R_1NR_2NH+R_3NX}$ for (3a-3b and 3d-3f) plus (1) & (2)
$R_1NH+R_2NH+R_3NH +\ldots \rightarrow \mathbf{R_1NX+R_2NX+R_3NX+\ldots}$ for (3a-3c and 3g-3j)

$$HOX + R_1NH \rightarrow \mathbf{R_1NX} + H_2O \quad (4a)$$

$$\mathbf{R_1NX} + MPO(Fe^{3+}) \rightarrow MPO\ (Fe^{4+}) + \mathbf{R_1NX^-} \quad (4b)$$

$$\mathbf{R_1NX^-} \rightarrow \mathbf{R_1N^-} + X^{\bullet} \quad (4c)$$

$$\mathbf{R_1NX^-} \rightarrow \mathbf{R_1N^{\bullet}} + X^- \quad (4d)$$

$$\mathbf{R_1N^{\bullet}} + R_2NH \rightarrow \mathbf{R_1NR_2NH} \quad (4e)$$

$$X^- + H_2O_2 + MPO\ (Fe^{4+}) \rightarrow MPO(Fe^{3+}) + HOX + {}^{\bullet}OH \quad (4f)$$

$$X^{\bullet} + R_2NH \rightarrow \mathbf{R_2NX} + H^+ \quad \text{(continue from 4c)} \quad (4g)$$

$$\mathbf{R_2NX} + MPO(Fe^{3+}) \rightarrow MPO\ (Fe^{4+}) + \mathbf{R_2NX^-} \quad (4h)$$

$$\mathbf{R_2NX^-} \rightarrow \mathbf{R_2N^-} + X^{\bullet} \quad (4i)$$

$$X^{\bullet} + R_3NH \rightarrow \mathbf{R_3NX\ldots} \rightarrow X^{\bullet} \text{ (multiple cycles)} \quad (4j)$$

Net: $H_2O_2 + R_1NH + R_2NH \rightarrow \mathbf{R_1NX} + \mathbf{R_1NR_2NH}$ for (4a-4b and 4d-4f)
$R_1NH + R_2NH + R_3NH +\ldots \rightarrow \mathbf{R_1NX} + \mathbf{R_2NX+ R_3NX+\ldots}$ for (4a-4c and 4g-4j)

Note that cycle 1 (reactions 3a-3b and 3d-3f, or reactions 4a-4b and 4d-4f) depends on $H_2O_2$, whereas cycle 2 (reactions 3a-3c and 3g-3j, or reactions 4a-4c and 4g-4j) is $H_2O_2$ independent due to the DET of $RN_1X$ into a $X^{\bullet}$ atom. The net effects of both reaction cycles are the alternations (damages) to at least two biological targets $R_1NH$ and $R_2NH$ *with no halogen consumption*. Similar reaction cycles can occur for other RHS such as $XO^{\bullet}$ and $X_2$. They are most effective (i.e., cycle 2 occurs), if the $X^{\bullet}$ atom is yielded, which was indeed observed for radiolysis of N-bromoimides but not for N-chloroimides [52,53]. Cycle 2 will be more significant with the formation of $X_2$ in reaction (1) (as $X_2 + e^- \rightarrow X^{\bullet} + X^-$ readily), and therefore will become dominant as the halogen level reaches above a threshold. Indeed, HOX was also observed to be in equilibrium with $X_2$ or BrCl in the MPO-$H_2O_2$-halide system under acidic conditions in the presence of excess $X^-$, as in phagosomes [31,32,43,44]. Cycles 1 and 2 are expected to continue for numerous cycles before they are terminated. Note that the peroxidase cycle [30-34] in catalyzing the reaction ($X^- + H_2O_2 \Leftrightarrow HOX + H_2O$), and the radical chain [52,53] in radiolysis of N-halogenated species ($R^\bullet + Br_2 \rightarrow RBr + Br^\bullet$; $Br^\bullet + RH \rightarrow H^+ + Br^- + R^\bullet$. Net: $Br_2 + RH \rightarrow RBr + Br^- + H^+$) as originally suggested by Goldfinger et al [59], have been proposed previously to enhance the reaction efficiency, but each of those reaction cycles occurs at the consumption of a halogen species. They are therefore distinct from the *$O_3$-depleting-like DET-based halogen reaction cycles* proposed here that have no cost of halogens (acting as 'catalysts'). Then a question arises: how would the halogen reaction cycles be terminated?

In atmospheric photochemistry of ozone loss, RHS is removed from the catalytic cycles once the reaction ends with HX or $XONO_2$ ("inactive reservoirs") [38,39]. In the phagosome (cell), however, this would not be the case as aqueous HCl can readily react with $OH^\bullet$/HOX to form $X^\bullet$/$X_2$ and the hydrolysis of $XONO_2$ will quickly occur to form $XO^\bullet$ and $NO_2$. It is worth noting that *the above proposed cyclic reactions are essentially enzyme-driven electron-transfer reactions*. Peroxynitrite has a half life of 1.9 s at pH 7.4 and its decomposition leads to a major product nitrate ($NO_3^-$) and a minor product of $CO_3^-$ or $NO_2$ with/without reaction with $CO_2$ [35,36]:

$$O_2^{\bullet-} + NO^\bullet \rightarrow ONOO^- + H^+ \Leftrightarrow ONOOH \rightarrow OH^\bullet + NO_2 \text{ (30\% yield)} \quad (5a)$$

$$\rightarrow NO_3^- + H^+ \text{(70\%)} \quad (5b)$$

$$O_2^{\bullet-} + NO^\bullet \rightarrow ONOO^- + CO_2 \Leftrightarrow ONOOCO_2^- \rightarrow NO_2 + CO_3^- \text{ (35\% yield)} \quad (5c)$$

$$\rightarrow NO_3^- + CO_2 \text{ (65\%)} \quad (5d)$$

Since the redox potential of $NO_2$ is close to that $OH^\bullet$, $ONOO^-$ is believed to act as an $OH^\bullet$-like oxidizing species in causing damaging effects [35,36]. Notably, $NO_3^-$ is known to be an extremely effective scavenger for weakly-bound electrons ($NO_3$ has an electron affinity of 3.94 eV, larger than any halogen) [50,55,56], with the measured reaction rate constant of as large as $1.2(\pm0.5)\times10^{13}$ $M^{-1}s^{-1}$ [56]. Reaction of $CO_3^-$ with electrons or $O_2^{\bullet-}$ is also expected (see Table 1). Thus, the halogen reaction cycles would be terminated if there is a significant amount of $NO_3^-$ or $CO_3^-$ (mainly $NO_3^-$) in the phagosome. Remarkably, *the reduction or removal of nitrogen species ($HNO_3$ as the ultimate product) is a critical requirement for the formation of the ozone hole (significant ozone loss) in the polar stratosphere [38,39].*

It follows that our proposed immune defence system fighting against pathogens is composed of three components: the $H_2O_2$ signaling pathway is the activator ('*trigger*'), the RHS reaction cycles the accelerator ('*amplifier*'), and the $NO^\bullet$ signaling pathway ($NO_3^-$) the deactivator ('*terminator*') of RHS reactions, as shown in Figure 2. The delicate balance between the two regulations, by $H_2O_2$ and $NO^\bullet$, is responsible for immunity that allows multicellular organisms to have adequate biological defenses to fight infection, disease, or other unwanted biological invasion, while having adequate tolerance to avoid allergy and autoimmune diseases/disorders. Note that as stable and small neutral signaling molecules that can diffuse easily across the membrane, both $H_2O_2$ and $NO^\bullet$ from other cellular organelles such as peroxisomes and mitochondria can also involve in the defense system [60].

### *2.2. Halogen Cyclic Reaction Rates*

It is striking to find the close similarity between the negative ion chemistry in the undisturbed stratosphere and troposphere of the Earth and that in phagosomes of the cell, as shown in Figure 3. Both are quite similar, involving identical species of $O_2^{\bullet-}$, $CO_2$, NO, $NO_2$, and $ONOO^-$ [61,62,19], except that $O_3$ is far more abundant in the stratosphere though its presence in phagosomes was also arguably observed [32]. For negative ion chemistry in the atmosphere, the primary ions are quickly converted into $CO_3^-(H_2O)_n$ ions on the order of $10^{-3}$ s by a series of reactions with $O_2$, $CO_2$, $O_3$ and $H_2O$ [61,62]. In the stratosphere, the $CO_3^-(H_2O)_n$ ions are converted relatively slowly into the more stable ions $NO_3^-(H_2O)_n$ or $NO_3^-(HNO_3)_m(H_2O)_n$ on the order of 1s by reaction with the trace gases NO, $NO_2$, $HNO_3$, and $N_2O_5$ [61,62]. For altitudes below 30 km, the hydrated $NO_3^-$ ions are the dominant and *terminal* species. The positive-ion chemistry is even simpler and the cluster ions $H_3O^+(H_2O)_n$ (n=3-6) are the sole primary ions since their formation is also rapid ($10^{-3}$ s). This is much shorter than the (positive) ion- (negative) ion recombination lifetime (i.e., the residence time of ions) of approximately 100-1000 s in the atmosphere [62].

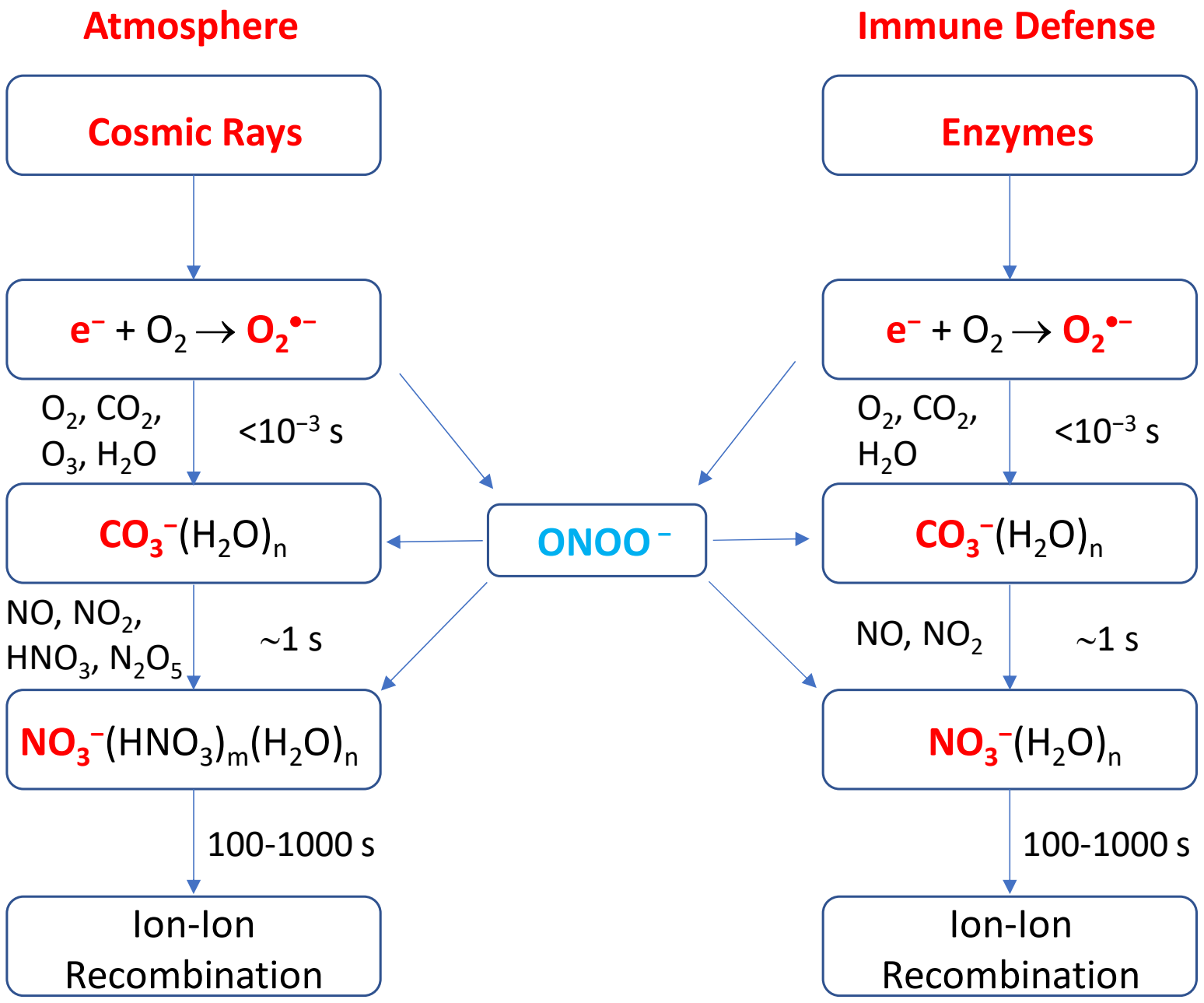

**Figure 3.** Schematic representations of the negative ion chemistry in the undisturbed stratosphere and troposphere of the Earth [61,62] and in the phagolysosome in the innate immune defense system in humans.

Looking into the rate constants in Table 1 and reactant abundances of all the cyclic reactions (1-5), we can see that reaction (1) or (3f) or (4f) is the rate-determining step, that is, the haloperoxidase catalyzed converting reaction of $X^-$ and $H_2O_2$ into HOX is the rate-limiting reaction. We further notice that $H_2O_2$ is stable, while the lifetime of negative ions including $Cl^-/Br^-/I^-$ and $NO_3^-$ is controlled by the recombination time with the abundant hydronium ions $H_3O^+(H_2O)_n$ (protons) in the *acidic* phagolysosome, which are long-lived, in contrast to hydronium in neutral solutions. It was indeed observed that the increase in IR-stimulated ROS/RNS generation persisted for *2–5 min* post-irradiation in most cell lines and for *~15 min* after irradiation in some cell lines [63]. Thus, the reaction period of RHS in the phagolysosome seems approximately 100-1000s, identical to that is the lower atmosphere. Given the $Cl^-$ and $Br^-$ concentrations in phagocytes are 120 mM (median) and 60 μM (median) [54] and the rate constants of $2.6\times10^4$-$3.9\times10^6$ $M^{-1}$ $s^{-1}$ and $3.0\times10^7$-$1.1\times10^8$ $M^{-1}$ $s^{-1}$ for the converting reactions with $H_2O_2$ at pH 5.0 respectively [64,65], we calculate that the overall Cl and Br cyclic reactions occur with respective rates of approximately 3120-$4.68\times10^5$ $s^{-1}$ and 1800-6600 $s^{-1}$. These give rise to *at least* $3.1\times10^5$-$4.7\times10^7$ and $1.8\times10^5$-$6.6\times10^5$ reaction cycles during the lifetime (≥100 s) of $Cl^-$ and $Br^-$ in the phagolysosome respectively. These estimated values are fairly close to the ozone depleting reaction cycles in the atmosphere [38]. These appear quite surprizing, but they are actually mainly determined by the levels of $Cl^-$ and $Br^-$ and the ion residence time. The latter is governed by the recombination with $H_3O^+(H_2O)_n$ [61,62], resulting in the apparently identical ion lifetime in both the stratosphere and the phagolysosome.

It must be noted that although the proposal of halogen reaction cycles in phagocytosis appears striking and novel, each of the reactions (1-5) is very well-established in the literature [15-19, 43-51]. Given the abundances and low reduction potentials [31, 32] of endogenous haloperoxidases (MPO/EPO/LPO) in phagosomes and the high resemblance [61,62,19] of the negative ion chemistry in phagosome to that in the atmosphere, there is every reason to deduce that the reaction cycles can be repeated thousands of times and amplify the RHS pathogen-killing (defense) effects by ~$10^5$ fold, analogous to the cyclic ozone-depleting reactions [38]. For the latter, RHS is known to be far more efficient than OH and NO at catalyzing the destruction of ozone, and among RHS, Br or I is much more efficient than Cl at catalyzing the cyclic reactions [18,19,38,39,45,46,49,50]

### *2.3. Significance of Halogen Cyclic Reactions*

This proposed mechanism has multifold significance:

1) The $H_2O_2$-initiated halogen cyclic reactions should greatly *amplify* the pathogen-killing effects of RHS in phagocytosis for innate and adaptive immune defenses. The proposed mechanism might lead to discoveries of highly effective antiviral drugs for effective treatment of various viruses such as HIV and COVID-19.
2) It also provides *a long-sought amplification mechanism* for reactive radicals produced by IR to generate the clinically required radiobiological effects in human body, as compared with background ROS.
3) It reveals that $H_2O_2$ as a signaling molecule plays *the key protective (pathogen-killing) role in immunological defenses* in many organisms from plants and animals to humans.
4) It also unravels *the crucial physiological role of $H_2O_2$* to account for the observed critical effect of enzymes SOD [8-15]. Once the SOD was impaired, the dysfunction of $H_2O_2$ as the protective molecule in the immune system resulted in and the animals died quickly.

### *2.4. Implications for the Pathology of COVID-19*

If people infected by an acute pathogen like SARS-CoV-2 that causes COVID-19 show mild and moderate symptoms and *receive no or improper treatment*, they might rapidly develop into severe cases of extreme respiratory burst. In the latter case, the halogen cyclic killing reactions may result in dysfunctions or disorders of closely interacting and communicating cell organelles, such as peroxisomes, mitochondria and lysosomes, and even severe damage to the host cells and tissues during autoimmune (over inflammation), by analogy with the severe springtime Antarctic ozone hole. The over toxicity caused by amplified RHS reactions is expected to result in unrepairable damage or mutations to cellular components, and the signaling pathways of $H_2O_2$ and $NO^{\bullet}$ may be impaired as well. Disorders in these organelles can cause serious damage to multiple organs such as lung, heart, liver, kidney and nervous systems, and respiratory failure, septic shock, multi-organ failure, or even deaths. *The best treatment should be in early or moderate stage or prevention*.

## 3. Halogenated Aromatic Drugs as Repurposed Drugs

Given our new understanding of the pathology of invading pathogens described above, we now discuss potential treatment of the severe virus COVID-19. Exogenous halogenated aromatic ring compounds as neutral small molecules that can freely diffuse across membrane will be potent anti-pathogen agents. Here we present a *new molecular mechanism of action* (MOA) of a class of halogenated aromatic drugs such as chloroquine and hydroxychloroquine (HCQ), which have been a current focus for experimental treatment of COVID-19 [66-68] and more potent di-amino, di-halogen aromatic ring molecules [termed *femtomedicine* compounds, FMDs], which have exhibited excellent targeted chemotherapy of multiple cancers [69-71]. *These drugs or candidates are promising for effective treatments of COVID-19, cancer and other diseases.*

For the MOA of HCQ and chloroquine, the readers are referred to a recent review [72]. As noted by the reviewers, most of the MOAs to explain the therapeutic and/or adverse effects were hypothesized based on *in vitro* studies, and the link between the proposed MOAs and the clinical efficacy and *in vivo* safety is yet to establish. The proposed molecular MOA of HCQ during autoimmunity include the following processes. Pharmacodynamic studies showed that these antimalarial drugs are lipophilic weak bases, easily cross plasma membranes, and preferentially accumulate in lysosomes, causing *increases* in the pH of lysosomes from 4 to 6 [73]. Elevation in pH causes inhibition of lysosomal activity including diminished proteolysis effect [74], decreased intracellular processing, glycosylation and secretion of proteins with many consequences [75]. These effects are believed to cause a decreased immune cell functioning such as chemotaxis, phagocytosis and $O_2^{\bullet-}$ production by neutrophils [76].

Another proposed MOA was that HCQ inhibits Toll-like receptor (TLR) signaling. This speculation might be traced back to a review [77]. The authors reviewed potent antiviral and antitumor properties of several imidazoquinolines as nucleic acid analogs, for which TLR7 was suggested to sense viral infection. Also, their unpublished data showed that two other immunomodulators, loxoribine and bropirimine, also activated immune cells through TLR7. Bropirimine (2-amin-5-bromo-6-phenyl-4(3)-pyrimidinone) is an immunomodulator that induces production of cytokines including IFN-$\alpha$. There has been interest in searching TLR-activating agents for clinical use. Takeda et al [77] predicted that new therapies utilizing the TLR-mediated innate immune activation would be developed to treat several disorders such as infection, cancer, and allergy. Because of some structural similarity to these imidazoquinolines, there have been speculations that TLR7 and TLR9 receptors involve in the MOA of HCQ.

Overall, available data suggest that HCQ and chloroquine impair or inhibit lysosomal and phagolysosome functions and subsequently immune activation. There are current efforts to identify exact molecular targets of HCQ within the lysosome, but convincing results and the identification of other molecular targets within the lysosomes have not yet reached. *The precise molecular mechanisms of anti-pathogen effects these drugs remain essentially unknown* [72].

### *3.1. New Molecular Mechanism of Action (MOA) of Halogenated Aromatic Drugs*

Here we propose a new molecular MOA of halogen-containing aromatic ring drugs represented by AXs, including HCQ and chloroquine, and a family of FMD halogenated compounds [69-71]. Here A is a quine ring system and X=Cl for HCQ and chloroquine, whereas A is an aromatic ring coupled to two $NH_2$ groups at ortho positions [A=B($NH_2$)$_2$] and X=$Cl_n$ or $Br_n$ or $I_n$ with n=1,2 for FMD compounds such as 4,5-dichloro/dibromo/diiodo-1,2-diaminobenzene and 4(5)-chloro/bromo/iodo-1,2-diaminobenzene [69,70]. We propose that these molecules are highly effective for DET reactions and their action is quite similar to the formation of endogenous RHS in the phagosome. It is long known that the DET reactions of halogenated aromatic organic molecules AXs with weakly-bound electrons in water are highly effective [18,19,45-51]. The exothermic energy for the DET reaction in water is $E_A$(X)-D(A-X)+$E_P$ [18,19,45-51], where $E_A$(X) and $E_P$ are defined earlier and D(A-X) is the known bond dissociation energy of C-X [D(C-Cl)=3.43 eV; D(C-Br)=2.81eV; D(C-I)=2.50 eV]. The thus estimated exothermic energies for the DET reactions of AXs to form $A^\bullet$ and $X^-$ (or $A^-$ and $X^\bullet$) are 1.48 (or 1.37), 1.85(or 1.99) and 1.86(or 2.3) eV, respectively for X=Cl, Br and I. These exothermicities are significantly larger than the reduction energies (1.09-1.16 eV) [31,32] of the redox couples (Compound I/native enzyme, $H_2O$) of enzymes MPO, EPO and LPO, indicating that the DET reactions of AXs must effectively take place in phagosomes, as shown in Figure 4. Again, the formation of $Br^\bullet$ or $I^\bullet$ atom (and $A^-$) is the favorable DET pathway for brominated and iodized AXs, in contrast to that for the DET of chlorinated AXs that favors the formation of $Cl^-$ (and $A^\bullet$).

Thus, similar to the DET reactions of endogenous halogenated organic compounds, effective DET reactions of AXs must occur with the enzymes such as MPO/EPO/LPO/SOD or NADPH which are weakly-bound electron donors:

$$e^- + AX \rightarrow AX^- \rightarrow (A^\bullet + X^-) \text{ or } (A^- + X^\bullet) \quad (6)$$

In addition, DET reactions of AXs with $O_2^{\bullet-}$ can also occur with or without the involvement of enzymes in the phagolysosome [78]:

$$AX + O_2^{\bullet-} \rightarrow (AO_2^\bullet + X^-) \text{ or } (AO_2^- + X^\bullet) \quad (7)$$

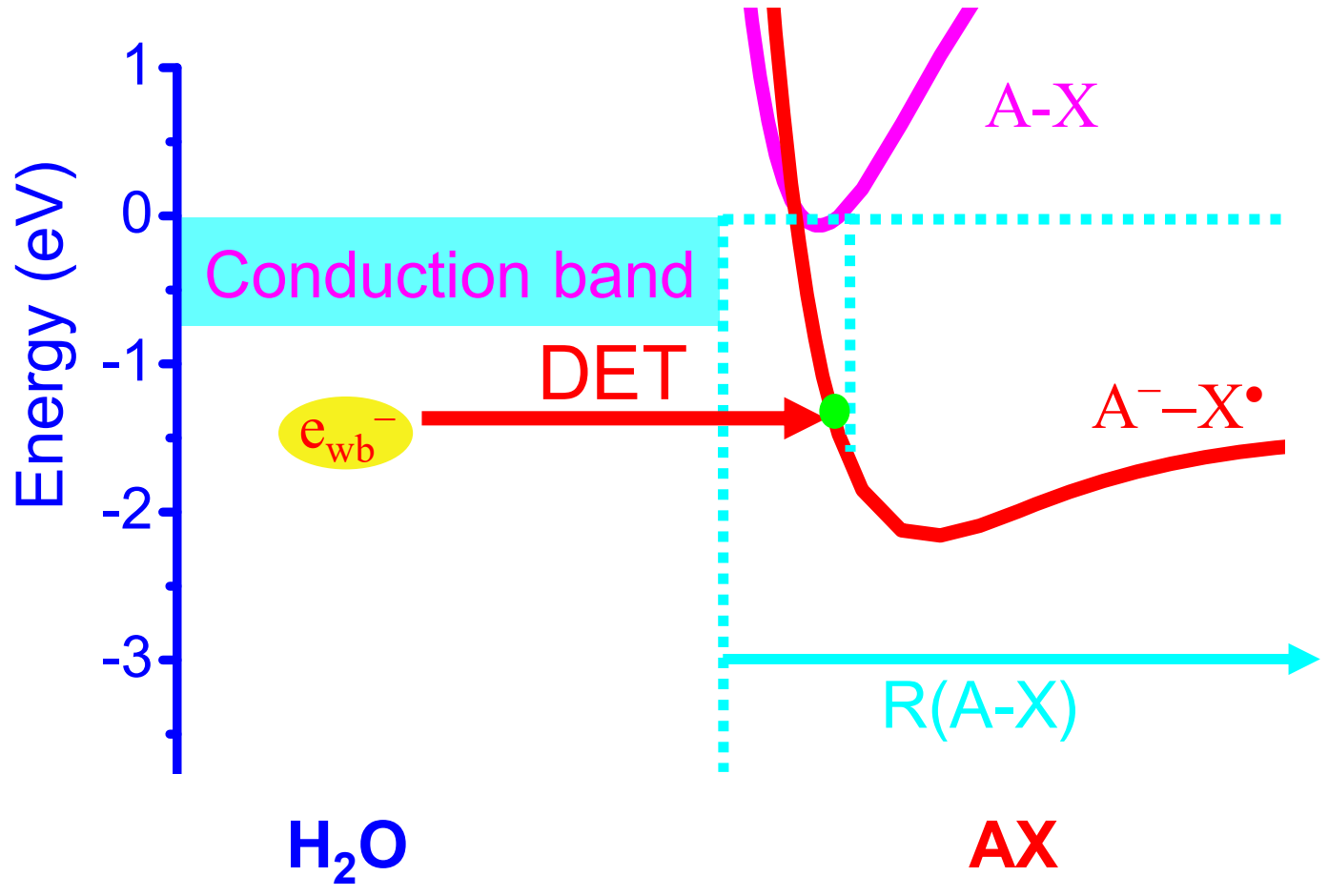

**Figure 4.** Schematic diagram for the dissociative electron transfer (DET) reaction [18,19,49,50] of a halogenated aromatic ring molecule AX with a weakly-bound electron ($e_{wb}^-$) in phagosomes, where $e_{wb}^-$ is readily donated from a variety of enzymes or antioxidants or $O_2^{\bullet-}$. For HCQ and chloroquine, A is a quine ring system and X=Cl, whereas A is an aromatic ring coupled to two $NH_2$ groups at ortho positions [A=B($NH_2$)$_2$] and X=$Cl_n$ or $Br_n$ or $I_n$ with n=1,2 for FMD compounds) [69,70].

Correspondingly, the halogen atom radicals $X^{\bullet}$ have particularly effective reaction cycles 2 to amplify the biological effects. The formed reactive radical $AO_2^{\bullet}$ can also directly cause damaging effects, e.g., by hydrogen abstraction from DNA/RNA, lipids or proteins, while $X^-$ like endogenous halogen ions can be converted into HOX, $OX^{\bullet}$, $X^{\bullet}$ and $X_2$ via the enzymes MPO/EPO/LPO. Like endogenous HOCl/HOBr/HOI, these exogenous RHS will have the amplifying cyclic reactions and cause significant anti-pathogen effects for immunological defence.

### *3.2. Potential New Therapeutics*

As observed for HCQ, these neutral small-molecule AXs including FMD compounds can easily cross plasma membranes to accumulate in lysosomes. We propose that the MOA of these halogenated drugs or candidates should be due to the *enhanced* (rather than *inhibited*) pathogen-killing effect induced by the reaction cycles of RHS derived from the DET of the compounds. *Indeed, the DET reaction of HCQ will lead to an increase in the $Cl^-$ yield, equivalent to a rise in pH in lysosomes, consistent with the observation for HCQ.* When there are invading pathogens in early or intermediate stages, pathogens are effectively killed in the formed phagolysosomes with excess RHS before autoimmune occurs. Since the DET reaction efficiencies are expected to increase in the order of Cl<Br<I for X in AXs [45,46,49,50], as we already observed for halopyrimidines (CldU, BrdU, IdU) [49,50], and more critically the DET reactions of Br- and I-containing aromatic ring molecules preferably produce $Br^{\bullet}$ or $I^{\bullet}$ atoms to activate cycle 2 reactions, the *in vivo* therapeutic efficacy should increase in the order of $ACl_n \ll ABr_n < AI_n$ (n=1,2).

As for anti-cancer activity of AXs, similar RHS are expected to be generated and to cause apoptosis and cancer cell killing. Moreover, since both $H_2O_2$ and NO are neutral small signaling molecules that can diffuse freely across the membrane and have high mobility, similar reaction cycles of RHS are expected to take place in other intracellular organelles such as mitochondria and peroxisomes which are electron (antioxidant)-rich. Indeed, recent *in vitro* and *in vivo* experimental results of multiple cancers have confirmed that the anti-tumor efficacies of the FMD compounds increases in the order of $ACl_2 \ll ABr_2 < AI_2$ [69,70]. Moreover, our unpublished data [79] by direct, real-time femtosecond laser spectroscopic measurements have confirmed the highly effective DET reactions of the FMD compounds.

Notably, the FMD compounds distinguish themselves from other drugs such as HCQ and chloroquine, bropirimine, BrdU and IdU, not only by yielding highly reactive $Br^{\bullet}$ or $I^{\bullet}$ atoms to initiate the powerful $H_2O_2$-independent reaction cycles but by a unique structural characteristic that *contains two $NH_2$ groups attached to the ring system at ortho positions.* This idea is based on a previous finding by Lu and Madey [16,47] that the presence of ammonia can cause giant enhancements in the DET reaction efficacies of CFCs adsorbed on ice surfaces by up to 30000 times, which was subsequently confirmed by Stähler et al. [51]. Although *in vivo* DET efficiencies may not be enhanced to this extent, there is every reason to expect that the DET reactions of the FMD compounds should be *at least 10 times* more effective than any other halogenated organic molecules without the two neighboring $NH_2$ groups in their structures. This was indeed observed by comparing the DET reactions of FMDs with BrdU and IdU [79]. BrdU is very similar to bropirimine in structure [77]. Thus, we predict that *the in vivo anti-pathogen effectiveness of FMD compounds with proper formulations will be at least 10 times higher than that of any of HCQ, chloroquine, bropirimine, BrdU and IdU.*

Of particular interesting is also the finding that in treated *normal* cells, a FMD compound at 100 μM caused an *increase* by about 40% of the reduced glutathione (GSH) level, which is an endogenous protective antioxidant [15,32], whereas the GSH level in human cervical (ME-180) cancer cells significantly dropped upon treated by the FMD and decreased to ~35% at 100 μM FMD [69]. This marked selective depletion of the GSH was consistent with the proposed preferential DET reaction to activate in cancer cells and with the observed selective cytotoxic effects of FMDs in vitro and in vivo [69-71]. These results showed that FMDs even have *a protective effect* on normal cells while killing abnormal cells and can be used as targeted chemotherapeutic agents [69,70].

Although no tests on antiviral activity of FMD compounds have been performed, our new studies on animal cancer models have shown that FMDs such as 4,5-dibromo-1,2-diaminobenzene and 4,5-diiodo-1,2-diaminobenzene at doses of ≤35 mg/kg (i.p) showed a superior efficacy in anti-cancer activity (leading to significant tumor shrinkage) to the clinical drugs such as cisplatin and gemcitabine at doses that showed significant toxic side effects. In contrast, the maximum tolerated dose (MTD) of 4,5-dibromo-1,2-diaminobenzene was measured to be 220±20 mg/kg (i.p), and no significant multiple dose toxicity in blood, liver, kidney, gut and spleen for 90 mg/kg/week for 4 weeks in mice (i.v) were observed. The oral MTD, which is usually at least 5 times higher, may be over 1000 mg/kg. *These results indicate that FMDs are essentially non-toxic but highly effective chemotherapeutic agents at used doses that can preferentially kill abnormal cells*. It is therefore useful for natural targeted chemotherapy of cancer and other diseases, which potentially include viral infections and autoimmune disorders. By contrast, due to the increased risk of developing retinopathy during treatment with HCQ, the current ophthalmology guidelines recommend a maximal daily dose of 5.0 mg/kg (actual body weight) per day for HCQ, while in cancer studies, high HCQ doses up to 1,200 mg (17-18 mg/kg) daily was prescribed [72,80]. *Given these data, FMD compounds are estimated to have a clinical safe dose at least 10 times higher than HCQ in humans. Thus, FMDs at a wide non-toxic dose range have high potential to outperform existing drugs such as chloroquine and HCQ that have known toxic side effects*.

Caly et al. [81] reported that the FDA-approved ivermectin, a protease inhibitor for the treatment of parasitic infections and for the replication of HIV and Dengue viruses, inhibits replication of SARS-CoV-2 in vitro, making it a possible candidate for COVID-19 drug repurposing research. Interestingly, pharmacological studies showed that the drug binds to glutamate-gated chloride channels (GluCls, present in neurons and myocytes) in the membranes of invertebrate nerve and muscle cells, causing increased permeability to $Cl^-$ ions, resulting in cellular hyper-polarization, followed by paralysis and death. The pathology proposed in this paper can well explain the MOA of this drug: the increased $Cl^-$ permeability will elevate the $Cl^-$ level and the associated RHS cyclic reactions to enhance the pathogen-killing effects.

Also, emerging studies to understand the pathogenesis of SARS-CoV-2 have just been published [82,83]. By analyzing human, non-human primate, and mouse single-cell RNA-sequencing (scRNA-seq) datasets and identifying ACE2 and TMPRSS2 co-expressing cells in lung type II pneumocytes, ileal absorptive enterocytes, and nasal goblet secretory cells, Ziegler et al [82] showed that ACE2 is *enhanced* in human nasal epithelia and lung tissue by interferon and influenza but not upregulated in mice. They proposed that SARS-CoV-2 could exploit species-specific interferon-driven upregulation of ACE2, a *tissue-protective mediator* during lung injury, to enhance infection. In contrast, by analysis of the transcriptional response to SARS-CoV-2 compared with other respiratory viruses, Blanco-Melo et al [83] found that cell and animal models of SARS-CoV-2 infections, as well as transcriptional and serum profiling of COVID-19 patients, consistently showed a unique and inappropriate inflammatory response, which was defined by *low levels* of Type I and III interferons juxtaposed to elevated chemokines and high expression of IL-6. The latter study therefore suggested that the *deficit* in innate antiviral defenses coupled with abundant inflammatory cytokine production is the defining and driving feature of COVID-19. It seems that a consistent pathogenesis of COVID-19 has not yet emerged. Further studies are obviously needed.

## 4. Recommendation for Experimental Treatment of COVID-19

Provided with the pathology of infection and molecular MOA of drugs proposed in this paper, drugs (candidates) like HCQ and FMDs will *enhance rather than inhibit* the endogenous pathogen-killing effect in the immune system in humans. *These halogenated aromatic drugs or candidates should be encouraged and recommended for experimental treatment of COVID-19, especially for preventive, mild and moderate stages, and probably treatment of developed chronic conditions.* In these applications, the halogen compounds should provide strong pathogen-killing effects. During autoimmune disorders, however, precaution should be taken if these drugs are to be used, because they will produce an excess number of exogenous RHS, much larger than that of endogenous RHS. In the latter case, using a NO (iNOS) inducer to enhance the production of the electron killer ($NO_3^-$/$CO_3^-$) neutralizing RHS might be helpful to the patients. This paper provides a plausible molecular mechanism that might help the clinical practice of the drugs. It may improve the clinical efficacy, control the toxicity within a tolerated range, and suggest the optimal timing/sequential administration or possible combination therapy. This might meet the current medical challenges faced in this global pandemic and save many lives across the globe. Furthermore, *this research also reveals that $H_2O_2$ is a necessary signaling molecules (angel) in the defense system fighting against pathogens and maintaining normal physiological functioning and signaling and its dysfunction can cause many diseases of organisms ranging from plants to humans.* The role of phagocytosis in both innate and adaptive immune defenses is probably underestimated under current context of immunology and virology. The new mechanistic insights into the immune defense might lead to discovery and development of more potent and/or targeted drugs for treatment of many diseases or conditions such as viral infections such as HIV and SARS-COV-2 (COVID-19), aging and cancer.

**Acknowledgments:** This work was supported in part by the Canadian Institutes of Health Research (a grant to Q.B.L, # 51771-10127) and Natural Science and Engineering Research Council of Canada (NSERC, a grant to Q.B.L, # 50503-10603). No other funding sources have been provided in the writing of this manuscript or the decision to submit if for publication. The author declares that there were issued patents or submitted patent applications associated with the data reported in this manuscript. Correspondence and requests for materials should be addressed to Q.B.L. (qblu@uwaterloo.ca).